\documentstyle[aps,prl]{revtex}

\def\l{\left}
\def\r{\right}
\def\lapl{\triangle}
\newcommand {\ex}[1]{ \exp \l\{ #1 \r\} }
\def\rf#1/ #2/ #3/ #4/ #5/{{#1}, {#2} {\bf #3}, {#4} ({#5})}

\begin{document}
\draft

\twocolumn[\hsize\textwidth\columnwidth\hsize\csname
@twocolumnfalse\endcsname

\title{On the Field Theoretical Approach to the Anomalous
Scaling in Turbulence}

\author{ Anton~V.~Runov }

\address{
Department of Theoretical Physics, St.Petersburg
University, Uljanovskaja~1, St.~Petersburg, Petrodvoretz,
198904, Russia}

\maketitle

 \begin{abstract}
Anomalous scaling problem in the stochastic Navier-Stokes
equation is treated in the framework of the field
theoretical approach, successfully applied earlier to the
Kraichnan rapid advection model. Two cases of the space
dimensions $d=2$ and $d\to\infty$, which allow essential
simplification of the calculations, are analyzed. The
presence of infinite set of the Galilean invariant
composite operators with {\em negative\/} critical
dimensions in the model discussed has been proved. It
allows, as well as for the Kraichnan model, to justify the
anomalous scaling of the structure functions. The explicit
expression for the junior operator of this set,
related to the square of energy dissipation operator, has
been found in the first order of the $\epsilon$~expansion.
Its critical dimension is strongly negative in the two
dimensional case and vanishes while $d\to\infty$.
\end{abstract}

]

\narrowtext

It is well known that the Kolmogorov-Obuchov~(KO) theory
fails to describe certain features of fully developed
turbulence. In particular, a good example is that of the
single time structure functions of velocity
\begin{equation}         \label{DefSF}
 S_n(r) \equiv \left\langle [v_L(r)-v_L(0)]^n
 \right\rangle, \qquad v_L=\frac{(\bbox{vr})}r.
\end{equation}
From the viewpoint of the KO~theory, their behavior in the
inertial range depends only on the mean energy dissipation
rate $\bar\varepsilon$
\begin{equation} \label{KOpred}
S_n(r) = C_n(\bar\varepsilon r)^{n/3},
\end{equation}
which leads to linearity of the corresponding exponent
in $n$. However, this is not consistent with experimental
results. This fact, often referred to as anomalous scaling,
is being actively discussed now. The problem is usually
treated in the framework of phenomenological
models~\cite{anom}, where it is related to large
fluctuations of the energy dissipation rate. The deviations
from the Kolmogorov scaling are described by additional
(anomalous) exponents $q_n$ as follows
\begin{equation}  \label{SFphen}
S_n(r) \simeq C_n(\bar\varepsilon r)^{n/3} (r/L)^{q_n},
\end{equation}
where $L$ is the integral scale of the flow. For
$n>3$ the experiment shows that $q_n$ is notably less then
zero.

The only statistical model allowing the calculation of
anomalous exponents is the Kraichnan rapid advection model,
introduced by Obukhov~\cite{Ob} and Kraichnan~\cite{Kr},
which attracts considerable interest being a simple model
for anomalous scaling investigation~\cite{anom_kr}. Recently
it was treated within the field theoretical approach, based
on the analysis of critical dimensions of so called
composite operators --- products of primary field and/or
their derivative at a single point~\cite{Adz98}.  This
approach provides one of the most efficient ways to
determine the universal scaling quantities, by ignoring all
inessential details, which is a great advantage in a
complex problem investigation. Particularly, it became
possible to calculate the anomalous exponents for the
Kraichnan model in the second order of the
$\epsilon$~expansion (the first order result in this case
is consistent with that obtained in~\cite{anom_kr}).
Another convenience of the approach in question is its
universality: it does not exploit the peculiarity of the
Kraichnan model and thus can be applied to a wide variety
of the statistical models.

In the present letter this approach is developed for the
stochastic Navier-Stokes~(SNS) equation in two cases of
space dimensions $d=2$ and $d\to\infty$, which allow
essential simplification of calculations. The obtained
results are found to be analogous to that for the Kraichnan
model. However, due to considerable technical complexity it
is only qualitative analysis that can be performed.

The starting point in the scaling investigation is the
dimensional analysis. We will distinguish momentum $d^k$
and frequency $d^\omega$ dimensions, so the dimension of an
arbitrary quantity $F$ is represented as
$[F]=\mbox{(length)}^{-d^k_F}\times
\mbox{(time)}^{-d^\omega_F}$. It is obvious that $d^k_k=
-d^k_r= 1$ and $d^\omega_\omega= -d^\omega_t= 1$. For the
viscosity $\nu$ one obtains $d^k_\nu=-2$, $d^\omega_\nu
=1$. The fully developed turbulence has two characteristic
lengths: the integral scale $L$, determined by the geometry
of the flow, and the dissipative length $l \equiv
\bar\varepsilon \,^{-1/4} \nu^{3/4}$. Thus, the natural
representation for an arbitrary function $S(r)$ is
\begin{equation} \label{dim}
  S(r)= \nu^{d^\omega_S}
    r^{-d_S} F\left(\frac rl, \frac rL\right).
\end{equation}
Here $d_S \equiv d^k_S + 2d^\omega_S$ is so called
canonical dimension of $S$, which plays an important role
in the field theoretical approach (as well as conventional
dimension for the static problems). Particularly, for the
structure functions~(\ref{DefSF}) $d^\omega_{S_n}=n$,
$d^k_{S_n}=-n$ and consequently $d_{S_n}=n$, so the
Eq.~(\ref{dim}) takes the form
\begin{equation}   \label{SFdim}
  S_n(r)=\nu ^n r^{-n} F_n(\frac rl, \frac rL).
\end{equation}
However, this representation is not fruitful for the fully
developed turbulence, because of the divergences of the
functions $F_n$ while the Reynolds number goes to infinity.
More precisely, one is interested in the structure
functions behavior in the inertial region $L\gg r\gg l$,
which corresponds to the limit $r/l \to\infty$,
$r/L \to 0$ in Eq.~(\ref{SFdim}). It should be noted
that there are two different problems: the asymptotical
behavior of the functions $F_n$ while the first argument
goes to infinity, and while the second one goes to zero.

The common way for treating the first problem is to isolate
the factor $(r/l)^{-\gamma_n}$ in Eq.~(\ref{SFdim})
\begin{equation}  \label{SFgamma}
 S_n(r)=\nu^n r^{-n} \left( \frac rl
 \right)^{-\gamma_n} \tilde F_n \left(\frac rl, \frac
 rL\right).
\end{equation}
If the choice of the exponents $\gamma_n$ can provide the
condition $\tilde F_n(\infty,r/L)=\mbox{const} \ne 0$, then
infrared scaling of structure functions $S_n(r)$ with {\em
fixed\/} parameter $r/L$
\begin{equation}
 S_n(r)=\lambda^{\Delta_n} S_n(r\lambda)
\end{equation}
will be described by their critical dimensions $\Delta_n
\equiv d_{S_n}+\gamma_n$.

In the framework of the KO~theory, functions $\tilde
F_n(r/L) \equiv \tilde F_n(\infty, r/L)$ are assumed to
have nonzero limit while $r/L \to 0$ and the exponents
$\gamma_n$ are determined from the condition of the
disappearing of $\nu$ dependence in the asymptotical
behavior of the structure functions in the inertial region
(this leads to the Eq.~(\ref{KOpred})).

On the other hand, critical dimensions of different
quantities can be calculated directly from the
corresponding stochastic equation by means of the
renormalization group~(RG) method. For the structure
functions~(\ref{DefSF}) the exact nonperturbative result
can be obtained~\cite{Obz}, which gives Kolmogorov's values
$\Delta_n=-n/3$. However, the RG~method does not require
the finiteness of the functions $\tilde F_n(r/L)$ in
Eq.~(\ref{SFgamma}) (the only condition is $l \ll L$). It
is obvious that the anomalous scaling appears in the case
of divergencies of the functions $\tilde F_n$ while $r/L
\to 0$.

Within the field theoretical approach, the asymptotical
behavior of the functions $\tilde F_n$ while $r/L \to 0$
can be analysed with the help of the operator product
expansion~(OPE), which allows to express the dependence
discussed in terms of critical dimensions of composite
operators. The detailed account for the RG and OPE methods
applied to SNS equation can be found in~\cite{Obz}, so here
we provide only necessary information. In general, OPE of
the discussed above functions $\tilde F_n$ has the form
\begin{equation} \label{Fn}
 \tilde F_n \propto \sum_{\Phi_k} C_n^{(k)}
 (r/L)^{\Delta_{\Phi_k}},
\end{equation}
where the coefficients $C_n^{(k)}$ are regular functions of
$r/L$ and thus can be treated as constants in the limit
$r/L\to0$. Summation in Eq.~(\ref{Fn}) is taken over all
composite operators $\Phi_k$, entering the OPE of $S_n$ (in
general, these are all possible operators allowed by the
symmetry), and $\Delta_{\Phi_k}$ are their critical
dimensions. The latter, unlike the critical dimensions
$\Delta_n$ of the structure functions, cannot be determined
from the simple dimensionality considerations. In this case
the application of the RG method is required, which allows
to calculate nontrivial critical dimensions
$\Delta_{\Phi_k}$ as well as coefficients $C_n^{(k)}$ in the
form of $\epsilon$~expansions.

One can see from Eq.~(\ref{Fn}) that the main contribution
to $F_n$ in the asymptotic region $r/L \to 0$ is given by
terms with negative $\Delta_k$, i.~e.\ by composite
operators with negative critical dimensions. These
operators were called ``dangerous'' in  \cite{Ad?}, because
the presence of such operators in the expansion~(\ref{Fn})
leads immediately to the divergences of the functions $F_n$
while $r/L\to0$ and, hence, to the anomalous scaling. As it
was pointed out in~\cite{gal-inv}, only Galilean invariant
composite operators enter into the OPE of invariant
objects, as the structure functions are. But it was not
until now that invariant dangerous operators have been
found in the SNS model. It is only the energy dissipation
operator
\begin{equation}
 \Phi_d=\frac12 \nu ( \partial_i \varphi_k +
 \partial_k \varphi_i )^2, \qquad
 \bbox{\varphi}=\bbox{v}-\langle \bbox{v} \rangle,
\end{equation}
that was known to have zero critical dimension~\cite{Obz}.

One of the main difficulties of the search of dangerous
operators in the SNS model originates from the mixing of
operators in renormalization. The matter is that an
arbitrary ultraviolet finite renormalized operator is a
linear combination of the primary one and those mixing in
renormalization. This results in splitting of the whole set
of operators into finite subsets --- ``families'', closed
under renormalization and containing operators of the same
canonical dimensions and symmetry properties~\cite{Obz}. As
a consequence, the renormalization of a composite operator
$\Phi_i$ is characterized by a renormalization matrix
$Z_{ij}$, rather than by a single constant as in the case
of ordinary quantities
\begin{equation}  \label{Z}
 \Phi_i=\sum_j Z_{ij} \Phi_j^R.
\end{equation}
Here the sum is taken over the full set of independent
operators from the corresponding operator family.
In fact, definite critical dimensions may be assigned to
the so called basis operators only, which diagonalize the
matrix $Z_{ij}$, and are determined by its eigenvectors.

Thus, the RG analysis of an operator family requires
calculating of the full renormalization matrix from
Eq.~(\ref{Z}). This provides possibility to determine all
basis operators and their critical dimensions. But the
number of independent operators in the SNS model grows
rapidly with the canonical dimension of the family, which
makes the analysis of senior families extremely laborious.

Particularly, we will be interested in the scalar Galilean
invariant operators as entering the OPE of the structure
functions. Canonical dimensions of the operators in
question are positive even integers. Up to the present
moment the families with $d_\Phi=2,4,6$ have been
examined~\cite{gal-inv,d6}. The mentioned above energy
dissipation operator $\Phi_d$ belongs to the family with
$d_\Phi=4$, and has the minimal critical dimension
$\Delta_{\Phi_d}=0$ among the families discussed.

This fact encourages us to consider the families,
containing powers of $\Phi_d$, to be the most likely
evidence of the presence of dangerous operators. The family
with $d_\Phi=8$, which contains the square of $\Phi_d$, was
examined in~\cite{Ver}. However, the analysis of the whole
family was not performed due to technical difficulties (the
family contains 12 independent operators). It was found
out only that $\Phi_d^2$ cannot be a basis operator
because of its mixing with the others: the fact which
breaks the equality
\begin{equation}   \label{Yk}
 \Delta[\Phi_d^n]=n\Delta[\Phi_d],
\end{equation}
proposed in~\cite{Yk}.

We succeeded in the analysis of the discussed above family
with canonical dimension 8 in two cases of space dimensions
$d=2$ and $d\to\infty$, where the essential simplification
of the problem occurs. Let us consider the first case in
more detail.

In this case the SNS equation for the velocity pulsation
field $\bbox{\varphi}$ can be reduced to the scalar
equation
  \begin{equation} \label{NSS}
\partial_t\lapl\phi = \nu\lapl^2\phi -
\partial_i\partial_m
(\varepsilon_{nm}\partial_n\phi\partial_i\phi) + f
  \end{equation}
for the stream function $\phi$ defined by the relation
$\varphi_i= \varepsilon_{ji}\partial_j \phi$
\cite{honk,my2}. Here $\epsilon_{ij}$ is the second-rank
antisymmetric tensor. Correlator of the random force $f$,
which imitate the interaction of velocity pulsations with
large-scale eddies, can be taken in the usual form
  \begin{mathletters} \label{COR}
  \begin{eqnarray}
\langle f(x,t) f(x',t')\rangle &&{}= \delta(t-t')\nonumber\\
  \times && \int\!\!
\frac{d{\bf k}}{(2\pi)^2} \, \ex{i {{\bf k} ({\bf x} - {\bf
x'})} } D({\bf k})\, ,
  \end{eqnarray}
  \begin{equation}
D({\bf k}) = D_0 k^{4-2\epsilon},
  \end{equation}
  \end{mathletters}
where $\epsilon>0$ is a formal small parameter of
RG~expansion. Its ``physical'' value $\epsilon=2$
corresponds to the case of energy pumping.
Correct RG~approach for this model, developed
in~\cite{honk,my2,Nal} gives the same results, as the
standard SNS~model.

Scalar formalism has at least two advantages: simpler
calculation of each diagram and less number of independent
operators. Let us see how the family of operators with
canonical dimension $d_\Phi=8$ can be treated within the
scalar approach of the model~(\ref{NSS}), (\ref{COR}). To
analyse an operator family one has to found the full set of
independent operators. Operators should be constructed from
the derivatives $\partial_i$, $\partial_t$ of field $\phi$
convoluted with tensors $\varepsilon_{ij}$, $\delta_{ij}$
and must necessarily be true scalars (an essential demand
as $\varepsilon_{ij}$ is a pseudo-tensor and $\phi$ is a
pseudo-scalar) and Galilean invariant. It can be shown that
the consideration may be restricted to the case of
operators irreducible to a full derivative of a junior
one~\cite{Obz}. Also, the specific symmetries of
two-dimensional space must be taken into account, leading
to identities of the type
$\varepsilon_{ij}\varepsilon_{kl}= \delta_{ik}\delta_{jl}-
\delta_{il}\delta_{jk}$.

As the result one obtains 7 independent operators:
 \begin{equation}  \label{bas}
  \begin{array}{l}
\Phi_1=(\triangle\phi)^4,     \\
\Phi_2=(\partial_i\partial_j\phi\:
\partial_i\partial_j\phi)^2, \\
\Phi_3=\partial_i\partial_j\phi\:
\partial_j\partial_k\phi\: \partial_k\partial_l\phi\:
\partial_l\partial_i\phi, \\
\Phi_4=\varepsilon_{il}\:
\triangle\partial_i\partial_j\phi\:
\partial_j\partial_k\phi\: \partial_k\partial_l\phi,
 \\
\Phi_5=\varepsilon_{il}\:
\nabla_t\partial_i\partial_j\phi\:
\partial_j\partial_k\phi\: \partial_k\partial_l\phi,
     \\
\Phi_6=(\triangle^2\phi)^2,     \\
\Phi_7=\nabla_t\partial_i\partial_j\phi\:
\nabla_t\partial_i\partial_j\phi.    \\
  \end{array}
 \end{equation}
Here $\nabla_t \equiv \partial_t + \varepsilon_{ji}
\partial_j \phi \partial_i$ is the substantial derivative.
The square of energy dissipation operator is expressed in
terms of operators (\ref{bas}) as follows
\begin{equation} \label{Phid}
 \Phi_d^2=-\Phi_1 +6\Phi_2 -4\Phi_3 .
\end{equation}

We have calculated full 7$\times$7 renormalization matrix
$Z_{ij}$ for the set of operators~(\ref{bas}) in the first
order of the $\epsilon$ expansion within the minimal
subtraction scheme.
As it was claimed in~\cite{Ver}, $\Phi_d^2$ is not a basis
operator and does not have any certain critical dimension.
But a dangerous operator still enters into the family in
question. It is
 \begin{eqnarray} \label{Phid'}
\Phi'_d &\cong& -\Phi_1 +6.03\Phi_2 -4.04\Phi_3 -0.58\Phi_4
  \nonumber  \\
&& {}+0.06\Phi_5 +0.10\Phi_6 -0.02\Phi_7
 \end{eqnarray}
with critical dimension
\begin{equation}
 \Delta_{\Phi'_d}\cong 8 -5.92\epsilon.
\end{equation}
At the ``physical'' point $\epsilon=2$ it becomes
$\Delta_{\Phi'_d} \cong -3.84$.

It is interesting to notice that the square of energy
dissipation operator~(\ref{Phid}) makes the major
contribution to the dangerous operator~$\Phi_d'$, while the
other operators of the family~(\ref{bas}) enter with small
numerical coefficients. The same is the case with the
critical dimension of~$\Phi_d'$: the major contribution
$8-6\epsilon$ is that of diagonal element of the
renormalization matrix, corresponding to the~$\Phi_d^2$,
and the admixtion correction  comprises only
$0.08\epsilon$. So the operator~(\ref{Phid'}) proves to be
dangerous due to the square of the energy dissipation
operator. However, it must be outlined that the discussed
critical dimension $\Delta_{\Phi'_d}$ considerably deviates
from $2 \Delta_{\Phi_d}=8-4\epsilon$ given by~(\ref{Yk}).

It is the second studied case of infinite dimensional
turbulence, where the exclusive role of the energy
dissipation operator becomes even more obvious. We have
shown that this is the case when the renormalization matrix
is block-triangular, so that the $\Phi_d^2$ operator is the
basis one. Its critical dimension
\begin{equation}
 \Delta_{\Phi_d^2} = 8-4\epsilon
\end{equation}
satisfies the Eq.~(\ref{Yk}) and equals to zero in the
``physical'' point as for the energy dissipation operator itself.
Moreover, we show within one-loop approximation that in the case
discussed the Eq.~(\ref{Yk}) proves to be valid for {\em all}
powers of $\Phi_d$~\cite{unpub}.

Let us show now that the existence of one dangerous
operator in the expansion~(\ref{Fn}) means that there is
an infinite set of dangerous operators. Indeed, by
combining inequality
\begin{equation}
 |S_n|^{1/n} \le |S_{n+1}|^{1/(n+1)},
\end{equation}
known from the probability theory with the
definition of anomalous exponents~(\ref{SFphen}), one
obtains inequality for even exponents $q_n$
\begin{equation} \label{ineq}
 \frac{q_n}n \le \frac{q_{n+2}}{n+2}, \qquad n=2m .
\end{equation}
Since $q_n$ are negative (which follows from both  the
experimental data and the presence of a dangerous operator
in the OPE of structure functions) the
inequality~(\ref{ineq}) implies that the absolute value of
$q_n$ grows more rapidly than n. Taking into account the
expansion~(\ref{Fn}) we see that only infinite set of
dangerous operators meet the last condition.

Thus, the results, obtained for the SNS model, proved to be
quite similar to those for the Kraichnan
model~\cite{Adz98}. In both cases the anomalous scaling of
the structure functions can be considered as a consequence
of the infinite set of dangerous operators in the
corresponding OPE, related to the powers of the energy
dissipation operator. Within the used approach, both models
exhibit the particularity of the infinite-dimensional case,
were the critical dimensions of all known dangerous
operators vanish (at least within the considered
approximations). In the Kraichnan model this is in
agreement with the disappearing of anomalous scaling and
intermittency while $d \to \infty$~\cite{Kr74}. Our results
make us expect that the analogous behavior takes place in
the SNS model also, which was first discussed in~\cite{Fr},
where the possible use of $1/d$~expansion for real
turbulence was pointed out.

However, the Kraichnan model contains at least two
principal simplifications. Firstly it is free of problems
connected with the mixing of operators: the critical
dimension of any power of the dissipation operator is
determined by the corresponding diagonal element of the
renormalization matrix (due to its triangularity). This
allows to calculate the critical dimensions of all
operators discussed in $\epsilon$~expansion~\cite{Adz98},
which seems impossible for the SNS model. Secondly, the
expansion~(\ref{Fn}) for any certain structure function
$S_n$ is finite in the Kraichnan model, which eliminates
the necessity of its summation. The anomalous exponent
$q_n$ in this case simply equals to the minimal $\Delta_k$
in Eq.~(\ref{Fn}). However there is no reason to expect
this in the SNS model, so the additional problem of the
summation of the expansion~(\ref{Fn}) arises. As a
consequence, the asymptotical behavior of the structure
functions can be more complicated than powerlike.

It becomes clear now that the developed here approach gives
qualitative account for anomalous scaling in the SNS model,
though, due to extreme difficulty of the problem it requires some
additional ideas for further investigations. In this context the
simple result for the infinite-dimensional case is of particular
interest. It implies the actuality of the detailed study of this
case as the starting point of $1/d$~expansion.

The author thanks Loran~Ts.~Adzhemyan for guidance and for many
stimulating discussions. Discussions with Nikolaj~V.~Antonov,
Alexander~N.~Vasil'ev and Mikhail~Yu.~Nalimov are also
acknowledged. This work was supported by the Russian Foundation
for Fundamental Research (Grant No.~96--02--17-033) and by the
Grant Center for Natural Sciences of the Russian State Committee
for Higher Education (Grants No~97-0-14.1-30 and No~M98-2.4K-567).


\begin{references}

\bibitem{anom} \rf M.~S.~Borgas/ Phys.~Fluids~A/ 4/ 2055/
1992/;
\rf C.~Meneveau and K.~R.~Sreenivasan/ Phys.~Rev.~A/ 41/
2246/ 1990/.

\bibitem{Ob} \rf A.~M.~Obuchov/ Izv.~Akad.~Nauk~SSSR,
Geogr.~Geofiz./ 13/ 58/ 1949/.

\bibitem{Kr} \rf R.~H.~Kraichnan/ Phys. Fluids/ 11/ 945/
1968/.

\bibitem{anom_kr} \rf M.~Chertkov, G.~Falkovich,
I.~Kolokolov, and V.~Lebedev/ Phys.~Rev.~E/ 52/ 4924/
1995/;
\rf K.~Gawedzki and A.~Kupiainen/ Phys.~Rev.~Lett./
75/ 3834/ 1995/.

\bibitem{Adz98} \rf L.~Ts.~Adzhemyan, N.~V.~Antonov, and
A.~N.~Vasil'ev/ Phys.~Rev.~E/ 58/ 1823/ 1998/;
\rf L.~Ts.~Adzhemyan and N.~V.~Antonov/ Phys.~Rev.~E/ 58/
7381/ 1998/.

\bibitem{Obz} L.~Ts.~Adzhemyan, N.~V.~Antonov, and
A.~N.~Vasil'ev, {\it The Field Theoretic Renormalization
Group in Fully Developed Turbulence,} (Gordon and
Breach, Amsterdam, 1999).

\bibitem{Ad?} \rf L.~Ts.~Adzhemyan, N.~V.~Antonov, and
A.~N.~Vasil'ev/ Sov.~Phys.~JETP/ 68/ 733/ 1989/.

\bibitem{gal-inv} \rf L.~Ts.~Adzhemyan, A.~N.~Vasil'ev, and
M.~Hnatich/ Teor.~Math.~Phys./ 74/ 115/ 1988/.

\bibitem{d6} \rf L.~Ts.~Adzhemyan, N.~V.~Antonov, and
T.~L.~Kim/ Teor.~Math.~Phys/ 100/ 1086/ 1994/.

\bibitem{Ver} \rf N.~V.~Antonov, S.~V.~Borisenok, and
V.~I.~Girina/ Teor.~Math.~Phys./ 106/ 75/ 1996/.

\bibitem{Yk} \rf V.~Yakhot, Z.-S.~She, and S.~A.~Orszag/
Phys.~Fluids~A/ 1/ 289/ 1989/.

\bibitem{honk} \rf J.~Honkonen/ Int.~J.~Mod.~Phys.~B/ 12/
1291/ 1998/.

\bibitem{my2} M.~B.~Orlov and A.~V.~Runov, Vestnik SPburg.,
Ser.~Fiz.~Khim. No~4(25), 120, (1997) [in Russian].

\bibitem{Nal} \rf J.~Honkonen and M.~Yu.~Nalimov/
Z.~Phys.~B/ 99/ 297/ 1996/.

\bibitem{unpub} To be published.

\bibitem{Kr74} \rf R.~H.~Kraichnan/ J.~Fluid~Mech./ 64/
737/ 1974/.

\bibitem{Fr} \rf J.-D.~Fournier, U.~Frisch, and H.~A.~Rose/
J.~Phys.~A: Math.~Gen./ 11/ 187/ 1978/.

\end{references}
\end{document}